\documentclass{article}

\usepackage{PRIMEarxiv}

\usepackage[utf8]{inputenc} 
\usepackage[T1]{fontenc}    
\usepackage{amsmath}
\usepackage{amsthm}
\usepackage{amssymb}
\usepackage{mathtools}
\usepackage{url}            
\usepackage{ltablex}
\usepackage{booktabs}       
\usepackage{amsfonts}       
\usepackage{nicefrac}       
\usepackage{microtype}      
\usepackage{lipsum}
\usepackage{fancyhdr}       
\usepackage{graphicx}       
\graphicspath{{media/}}     
\usepackage{wrapfig}
\usepackage{times}
\usepackage{latexsym}
\usepackage{bm}
\usepackage{tikz}
\usepackage{siunitx}
\usepackage{array}
\usepackage{multirow}
\usepackage{makecell}
\usepackage{verbatim}
\usepackage{color}
\usepackage{float}
\usepackage{arydshln}
\usepackage{longtable}
\usepackage{tabularx}
\usepackage[round,sort]{natbib}
\usepackage{hyperref}       
\usepackage{cleveref}
\usepackage{subfigure}
\usepackage{algorithmic}
\usepackage{algorithm}

\definecolor{mydarkblue}{rgb}{0,0.08,0.45}
\hypersetup{ %
	colorlinks=true,
	linkcolor=mydarkblue,
	citecolor=mydarkblue,
	filecolor=mydarkblue,
	urlcolor=mydarkblue}

\begin{document}
\title{USE: Dynamic User Modeling  with \\ Stateful Sequence Models}

\author{
\textbf{Zhihan Zhou}$^{\dagger}$\thanks{Equal contribution}
\qquad
\textbf{Qixiang Fang}$^{\ddag}$\footnotemark[1]
\qquad
\textbf{Leonardo Neves}$^{\S}$
\qquad
\textbf{Francesco Barbieri}$^{\S}$ 
\\
\textbf{Yozen Liu}$^{\S}$ 
\qquad
\textbf{Han Liu}$^{\dagger}$
\qquad
\textbf{Maarten W. Bos}$^{\S}$ 
\qquad
\textbf{Ron Dotsch}$^{\S}$ 
\\
$^\dagger$Northwestern University, USA  \:
$^\ddag$Utrecht University, Netherlands \: 
$^\S$Snap Inc, USA \\ 
{\footnotesize 
   \href{mailto:zhihanzhou@u.northwestern.edu}{\texttt{zhihanzhou@u.northwestern.edu}},   
   \href{mailto:q.fang@uu.nl}{\texttt{q.fang@uu.nl}}
   }
   \\
{\footnotesize 
    \href{mailto:lneves@snapchat.com}{\texttt{\{lneves}}, \href{mailto:yliu2@snapchat.com}{\texttt{yliu2\}@snapchat.com}},   
   \href{mailto:hanliu@northwestern.edu}{\texttt{hanliu@northwestern.edu}}}  \\
{\footnotesize
   \href{mailto:maarten.w.bos@gmail.com}{\texttt{maarten.w.bos@gmail.com}},
   \href{mailto:rdotsch@gmail.com}{\texttt{rdotsch@gmail.com}}} 
}

\maketitle

\begin{abstract}
User embeddings play a crucial role in user engagement forecasting and personalized services. Recent advances in sequence modeling have sparked interest in learning user embeddings from behavioral data. Yet behavior-based user embedding learning faces the unique challenge of dynamic user modeling.
As users continuously interact with the apps, user embeddings should be periodically updated to account for users' recent and long-term behavior patterns.
Existing methods highly rely on stateless sequence models that lack memory of historical behavior. They have to either discard historical data and use only the most recent data or reprocess the old and new data jointly. Both cases incur substantial computational overhead. 
To address this limitation, we introduce the \textbf{U}ser \textbf{S}tateful \textbf{E}mbedding (USE). USE generates user embeddings and reflects users’ evolving behaviors without the need for exhaustive reprocessing by storing previous model states and revisiting them in the future. 
Furthermore, we introduce a novel training objective named future $W$-behavior prediction to transcend the limitations of next-token prediction by forecasting a broader horizon of upcoming user behaviors. By combining it with the Same User Prediction, a contrastive learning-based objective that predicts whether different segments of behavior sequences belong to the same user, we further improve the embeddings' distinctiveness and representativeness. 
We conducted experiments on 8 downstream tasks using Snapchat users' behavioral logs in both static (i.e., fixed user behavior sequences) and dynamic (i.e., periodically updated user behavior sequences) settings. We demonstrate USE's superior performance over established baselines. The results underscore USE's effectiveness and efficiency in integrating historical and recent user behavior sequences into user embeddings in dynamic user modeling.

\end{abstract}

\section{Introduction}

The era of digital transformation has ushered in an unprecedented emphasis on personalization, primarily driven by the ability to understand and predict user behavior. In this context, user embeddings – numerical vector representations of user characteristics, behavioral patterns, and preferences – have become indispensable. These embeddings are central to a myriad of applications, from recommendation systems to targeted advertising \citep{chen_predictive_2018, wu_ptum_2020, modell_graph_2021}, and their effectiveness directly influences user experience and engagement. In the present work, we study general-purpose user embeddings that can be directly used for various downstream tasks without fine-tuning the upstream user embedding model, as opposed to task-specific user embeddings \citep{fan2019graph,waller_generalists_2019,zheng2017joint,liu2010personalized}. 

\begin{wrapfigure}{l}{0.5\textwidth}
    \centering
    \begin{tikzpicture}
    \node[inner sep=0] {\includegraphics[width=0.5\columnwidth, trim={12.5cm 7.2cm 23cm 3.5cm}, clip]{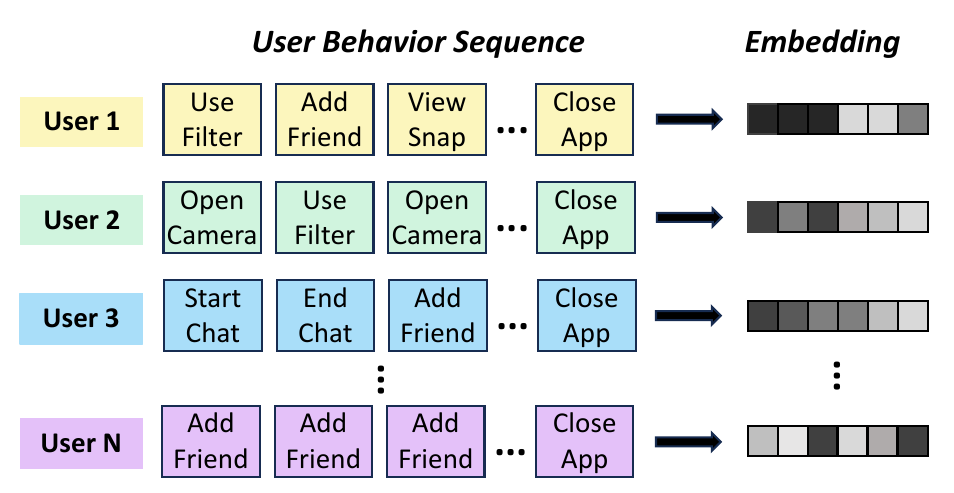}};
		\end{tikzpicture}
		\caption{Illustration of behavior-based user embedding. We aim to represent each user with a fixed-size numerical vector based on their behavior sequences.}
    \label{fig:user_embedding}
\end{wrapfigure}

User embeddings can be calculated from various data sources such as demographic data (e.g., age, gender, etc.) and user-created content (e.g., photos and messages). In this work, we focus on behavior-based user embedding models (see Figure \ref{fig:user_embedding} as an example), which take solely behavior sequences (e.g., open\_app -> open\_camera -> apply\_filter -> close\_app) as input to compute user embeddings. Such sequences reflect the actions users take in an app.  A good user embedding should reflect both their recent behavior as well as more long-term and potentially recurring behavior. Thus, as users continuously interact with an app, their embeddings should be periodically updated in response to their newly extended behavior sequences. 
This requires that user embedding models be efficiently updated with new behavioral data without sacrificing information about past user behaviors.

Existing techniques for behavior-based user embeddings  \citep{zhang_general-purpose_2020, pancha_pinnerformer_2022, chu_simcurl_2022} predominantly rely on stateless models (e.g., Transformers \citep{transformer}), which generate outputs purely based on current inputs without the memorization of historical inputs going back further than the context window of the current input. While these methods are powerful in capturing complex patterns in user behavior, they exhibit significant limitations in dynamic environments due to their inability to efficiently incorporate both historical and new data when updating user embeddings. When confronted with new behavior sequences, these models face a trade-off: either disregard historical data for efficiency, leading to a loss of valuable long-term behavioral insights or compute embeddings from scratch by processing all available data, incurring substantial computational costs and delays. In the first case, a possible strategy to include historical information is pooling the old embeddings with new embeddings based on the incoming data. However, our empirical analysis reveals that computing new embeddings without conditioning on historical user data can still result in substantial information loss, as detailed in Table \ref{tb:next_period_prediction} and \ref{tb:user_reid}. This problem is even more challenging in traffic-intensive apps where often hundreds of events are generated by each user every day. 

To address this challenge, we introduce USE, which stands for \textbf{U}ser \textbf{S}tateful \textbf{E}mbeddings, an approach that can efficiently produce user embeddings equivalent to explicitly incorporating all available data (including historical information) as model input, while maintaining constant computational costs regardless of the amount of user historical data.
Specifically, USE retains a state of past behaviors for each user; as new behavior data comes in, it efficiently computes new user embeddings by updating the previous user state, without the need to explicitly use all available data as model input.

To implement USE, we consider several aspects. The first is model architecture. We achieve statefulness by adopting the Retentive Network (RetNet) \citep{retnet}. RetNet was originally designed for natural language processing. It can be trained in parallel like Transformers while making inferences sequentially like Recurrent Neural Networks (RNN). The Transformer-like architecture ensures training scalability and user representation capability, while the sequential inference allows efficient updates of user embeddings, making RetNet an ideal choice for dynamic user modeling.

Second, we consider the training target. Next-token prediction has become the default pre-training target for almost all existing large language models (LLM) \citep{gpt2, retnet}. This training objective closely matches the natural language generation process, yet it is less suitable for user embedding learning. Given the stochastic nature of user behavior and the absence of consistent syntax and grammar in user behavior sequences, accurately predicting the exact next user behavior is not only less feasible but also likely detrimental to the model's ability to capture longer term user interests. 
Thus, we relax the order component in traditional next-token prediction and introduce a unique training objective named \textit{Future $W$-Behaviors Prediction} (FBP). 
Instead of predicting the exact next behavior, FBP trains the model to predict the presence of all the behaviors in the future $W$ user behaviors, where a larger $W$ emphasizes longer-term user engagements and a smaller $W$ emphasizes shorter-term user engagements. 
FBP thus prevents the model from overfitting to the idiosyncrasies in user behavior sequences, while allowing the embeddings to capture longer-term user intentions.


Third, we consider the encoding of both user states and user traits. Ideal behavioral user embeddings should encode both the state component of a user (what the user might do based on the current state, as captured with future $W$-behaviors prediction) and the trait (habitual) component of user behavior (what the user tends to do most of the time).  To this end, we incorporate the Same User Prediction (SUP), a contrastive learning-based training objective that predicts whether different segments of behavior sequences belong to the same user. This way, the model 1) is forced to learn to represent stable user features (traits) in addition to representing momentary behaviors (states), and 2) improves the distinctiveness and representational capability of the embeddings for individual users. 

To validate our approach, we conduct extensive empirical analyses using behavioral sequences from users of Snapchat, a popular multimedia instant messaging platform, with about 400 million daily active users as of the second quarter of 2023~\citep{snapchat_2023}. Overall, our analyses demonstrate the superiority of USE over strong baselines in both static and dynamic settings. Specifically, by evaluating USE on $8$ downstream tasks, we illustrate the effectiveness of the combined future $W$-behavior prediction and same user prediction training objectives over typical NLP training objectives in the context of user modeling. 
Furthermore, by comparing user embeddings generated using identical models but with different updating strategies, we demonstrate both the effectiveness and efficiency of USE in seamlessly integrating historical and recent user behavior data into user embeddings.
Compared to other approaches, USE not only substantially reduces computational costs but also consistently delivers higher-quality user embeddings across nearly all evaluation settings.

Our contributions are threefold: 1) we introduce stateful 
 user embeddings, 2) we develop novel training objectives to enhance these embeddings, and 3) we empirically demonstrate the superiority of USE over existing methods. This work not only advances user modeling research but also sets a new standard for efficient and effective personalization in dynamic online environments.

\begin{figure*}[t]
    \centering
    \begin{tikzpicture}
        \draw (0,0 ) node[inner sep=0] {\includegraphics[width=1\columnwidth, trim={1.2cm 9cm 16.5cm 1.5cm}, clip]{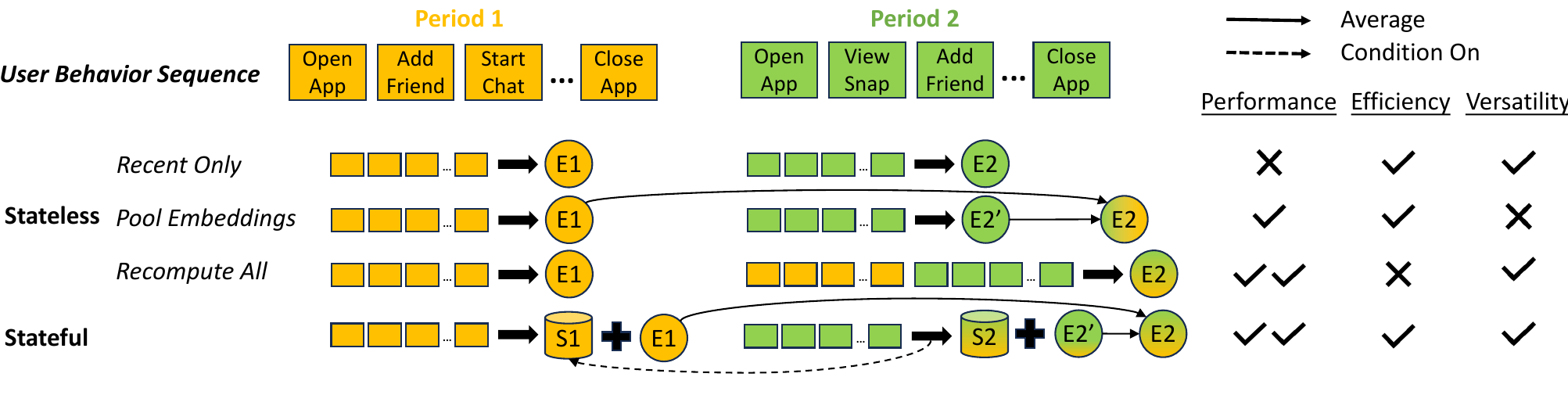}};
    \end{tikzpicture}
    \caption{Visualization of dynamic user modeling. 
    The empty rectangles indicate user actions; the round shapes indicate user embedding; the drum shapes indicate user states.}
    \label{fig:embedding}
\end{figure*}

\section{Preliminaries}

\subsection{Behavior-based User Modeling} 

User embeddings (fixed-size numerical vectors representing user characteristics and preferences) play a crucial role in personalization systems. These systems encompass a range of applications, including user understanding, detection of malicious users, friend suggestions, and item recommendations. Various methodologies have been developed to derive user embeddings from different types of user data \citep{modell_graph_2021, fan2019graph,waller_generalists_2019,zheng2017joint,liu2010personalized}.

This work focuses on behavior-based user modeling. Our objective is to compute user embeddings based on their behavior sequences, while deliberately avoiding the use of demographic data (e.g., age, race, nationality, and gender) or user-generated content (e.g., posts and messages), which are commonly used in the field of user modeling. 
Additionally, deriving user embeddings exclusively from natural interactions with the app reduces the need for active user input, like filling out surveys, thereby minimizing user burden and enhancing user experience.

Let us define $\mathcal{B} = \{b_1, b_2, \ldots, b_B\}$ as the set of $B$ unique user behaviors, where each behavior $\{b_i\}_{i=1}^B$ represents a distinct type of user interaction with the app (for example, opening the app or sending a message). Our model solely relies on the type of these user interactions. Let $\bm{x}^k = [x_1^k,x_2^k,\ldots,x_n^k]$ denote the behavior sequence of the $k$-th user $\bm{u}^k$, with each $x_i^k \in \mathcal{B}$. The behavior-based user models take $\bm{x}^k$ as input to generate a fixed-size vector $\bm{e}^k$, representing the embedding of user $\bm{u}^k$.

\subsection{Stateless and Stateful User Modeling}

A key challenge in behavior-based user modeling lies in the dynamic nature of user behavior sequences. As users interact with the app over time, their behavior sequences expand with new entries. To reflect the latest user behaviors accurately, it is essential to periodically update user embeddings.

For notation simplicity, let $\bm{x}_{p_0} = [x_1,x_2,\ldots,x_{p_0}]$ be the initial behavior sequence for user $\bm{u}$, where $p_0$ denotes the index of the last behavior up to the first computation period (period $0$). The user's embedding $\bm{e}_{p_0}$ is derived from $\bm{x}_{p_0}$. After some period (e.g., a week), a new behavior sequence $\bm{x}_{p_1} = [x_{p_0+1},\ldots,x_{p_1}]$ is generated, necessitating the recomputation of the user embedding, now based on both $\bm{x}_{p_0}$ and $\bm{x}_{p_1}$. This process repeats with each new period, accumulating more behavior data for embedding calculations.

In this dynamic setting, a \texttt{stateless} model (such as Transformer \citep{transformer} and Convolutional Neural Networks \citep{resnet}), which computes outputs based solely on current input sequences, can adopt one of three strategies for periodic embedding updating: \textbf{Recent Only}, \textbf{Pool Embeddings}, and \textbf{Recompute All} (visualized in Figure \ref{fig:embedding}).
\textbf{Recent Only} discards historical data, relying only on the latest user behavior sequences for embedding generation. \textbf{Pool Embeddings} enhances the first strategy by combining new embeddings with previously calculated ones. \textbf{Recompute All} always uses the entire user behavior sequence as input to compute embeddings at each period. While Recent Only and Pool Embeddings are computationally efficient, they disregard historical behavior sequences when computing outputs for sequences in the new period and may thus sacrifice effectiveness. Recompute All, while utilizing complete user history, incurs high computational costs.

The concept of a \texttt{stateful} user model is meant to address these limitations of stateless models. By maintaining and utilizing historical intermediate results in future computations, a stateful model achieves high computational efficiency without losing historical information. At period $p_{i-1}$, the stateful model calculates the user embedding ${\bm{e}_{p_{i-1}}}$ and produces a state ${\bm{s}_{p_{i-1}}}$—the model's memory of the user's relevant history. In the subsequent period $p_i$, the model processes the new behavior sequence $x_{p_{i}}=[x_{p_{i-1}+1},\ldots,x_{p_i}]$ alongside the last state ${\bm{s}_{p_{i-1}}}$, thus generating embeddings that encompass the entire user behavior history efficiently and effectively.

\section{Method}

\subsection{Model Architecture}
\label{subsec:method_architecture}
Theoretically, \textsc{USE} can be implemented with any model architecture that enables recurrent inference. 
We choose the implementation of RetNet~\citep{retnet} for its demonstrated effectiveness and efficiency over earlier methods in natural language processing research. The key difference introduced by RetNet is replacing the Attention operation with Retention, a operation than can be equivalently formatted in parallel and recurrent ways. Retention's parallel form empowers RetNet with Transformer-level scalability and representation power, while its recurrent form enables the modeling of states and efficient inference.
In this section, we focus on the core idea behind our implementation of stateful embeddings with RetNet. Please refer to the original paper \citet{retnet} for more technical details.


Let $\bm{x} = [x_1, x_2, \ldots, x_n]$ define an input sequence of length $n$ and $q_i$, $k_i$, and $v_i$ respectively define the query, key, and value matrices of $x_i$ at a Transformer/RetNet layer. Let $o_n^a$ and $o_n^r$ define the output of $x_n$ after respectively the attention and retention calculation, where the attention operation is defined as follows:
\begin{equation}
    o_n^a = \texttt{softmax} ([q_nk_1^\top, q_nk_2^\top, \ldots, q_nk_n^\top]) \cdot [v_1, v_2, \ldots, v_n]
\end{equation}
Because of the non-linear \texttt{softmax} function, to calculate the $n$-th output $o_n$, we need to perform the dot product between $q_n$ and all the previous $k$ before applying \texttt{softmax}. This leads to a computational complexity of $O(n)$. The retention operation, however, removes the \texttt{softmax} function, so that computation between $q$, $k$, and $v$ can be reordered by performing $k^\top v$ first, enabling the following definition of the retention operation:
\begin{equation}
\label{eq:retention}
\begin{split}
o_n^r & = \sum_{m=1}^n \gamma^{n-m} q_nk^\top_mv_m  = q_n \sum_{m=1}^n \gamma^{n-m} k^\top_mv_m \\ & = q_n(\gamma^{n-1} k_1^\top v_1 + \gamma^{n-2} k_2^\top v_2 + \cdots +\gamma^{0} k_n^\top v_n)
\end{split}
\end{equation}
where $\gamma$ is a hyperparameter between $0$ and $1$ that explicitly reduces the importance of distant tokens in current output. Let $s_1 = k_1^\top v_1$ and $s_n = \gamma s_{n-1} + k_n^\top v_n$, we have $(\gamma^{n-1} k_1v_1 + \gamma^{n-2} k_2v_2 + \cdots +\gamma^{0} k_nv_n) = s_n$. Equation \ref{eq:retention} can be written as $o_n^r = q_ns_n = q_n(\gamma s_{n-1} + k_n^\top v_n)$. In other words, current output $o_n^r$ only depends on current query $q_n$, $k_n$, $v_n$, and the latest state $s_{n-1}$, leading to computational complexity of $O(1)$. More importantly, computing embeddings in this way results in identical outputs as feeding the entire behavior sequence as input. 

This operation can be further extended to chunk-wise recurrent inference. When the latest state $s_{n-1}$ is pre-computed on input $[x_1, x_2, \ldots, x_{n-1}]$, the computational cost on the input sequence $[x_n, x_{n+1}, \ldots, x_{n+m}]$ only depends on $m$, regardless of the historical behavior length $n$. Note that, in the chunk-wise recurrent inference, the output of each $\{x_i\}_{i=n}^m$ is performed in parallel. Therefore, compared to purely recurrent models such as RNNs, it is much more efficient without sequential dependency within $[x_n, x_{n+1}, \ldots, x_{n+m}]$.

In dynamic user modeling, we start with a user behavior sequence $\bm{x}_{p_0} = [x_1,x_2,\ldots,x_{p_0}]$ for user $\bm{u}$. We initialize the first state $\bm{s}_0$ as an all-zero matrix and perform a chunk-wise recurrent forward pass to get the last hidden states of each input behavior $\bm{h}_{p_0} = [h_1,h_2,\ldots,h_{p_0}]$ as well as new state $\bm{s}_{p_0}$. The user embedding is calculated as $\bm{e}_{p_0} = \frac{1}{p_0} \sum_{i=1}^{p_0} h_i$. After a certain period, when new behavior sequence $\bm{x}_{p_1} = [x_{p_0+1},\ldots,x_{p_1}]$ is available, we perform another chunk-wise recurrent forward pass with $\bm{x}_{p_1}$ and $\bm{s}_{p_0}$ as input and obtain $\bm{h}_{p_1} = [h_{p_0+1},\ldots,h_{p_1}]$ and $\bm{s}_{p_1}$. The user embedding is then calculated as $\bm{e}_{p_1} = \frac{p_0}{p_1} \bm{e}_{p_0} + \frac{p_1-p_0}{p_1} \bm{e}_{p_1^*}$, where $\bm{e}_{p_1^*} = \frac{1}{p_1 - p_0} \sum_{i={p_0}+1}^{p_1} h_i$ is the average of new hidden states.

\subsection{Training Objectives}
\label{subsec:method_objective}
We aim to train a user model that can predict users' future engagements with the app and discriminate against different users, such that the user model can handle a wider range of downstream tasks. Specifically, we reason that user behavior forecasting may allow for more accurate item/ad recommendations and early detection of bad actors (e.g., users that violate rules of operations), while user discrimination may empower better personalization and user re-identification. With such design principles, we introduce two model training objectives: Future $W$-Behavior Prediction (FBP) and Same User Prediction (SUP).

\subsubsection{Future $W$-Behavior Prediction}
In user modeling, encoding users' long-term engagements is crucial. Typically, causal language modeling (CLM) is employed as a training objective, which focuses on predicting the immediate next user behavior based on a given behavior sequence. While CLM is a prevalent pre-training objective in natural language processing, it might not be optimal for behavior-based user models. Unlike natural languages with strict syntax and grammar, user behaviors are much more random and noisy. Forcing a model to predict the exact order of next user behaviors may lead to overfitting to the idiosyncrasies in the data and thus compromise effective user representation learning.

Future $W$-Behavior Prediction is designed to overcome this by relaxing the order constraint. In this approach, given a user behavior sequence and a specified future window size $W$, the objective is to predict the probability of each behavior occurring within the user's next $W$ actions. This training target, in contrast to CLM, prioritizes understanding a user's longer-term future interests over precisely predicting the sequence of their imminent behaviors. This focus more closely aligns with the goal of user modeling.

Let $N$ define the number of behaviors of interest. The label of future $W$-behavior prediction is a $N$-length binary vector, where each position indicates the presence of each unique behavior in the next $W$ user behaviors. We add a prediction layer that takes the last hidden state as input and makes $N$ predictions based on it. 
Given a sequence $\bm{x}^k = [x_1^k,x_2^k,\ldots,x_T^k]$ of length $T$, we predict the presence of every event of interest in $[x_{i+1}^k,\ldots,x_{i+W}^k]$ at each $x_i^k, 0 < i < T-W$.\footnote{In preliminary experiments, we explored different prediction strategies, including at the end of the sequence $x_{T-W}^k$, throughout the entire sequence, only in later parts, or at intervals. We found that training at the end of the sequence underperformed, while other methods yielded comparable results. Thus, we settled on the most intuitive approach, namely training at all behaviors in $[x_1^k, \ldots, x_{T-W}^k]$.} Let $\hat{y}^k_{i,n}$ and $y^k_{i,n}$ respectively define the predicted and actual presence of the $n$-th behavior of interest at the $i$-th input behavior $\bm{x}^k$, the FBP loss on $\bm{x}^k$ is defined as:

\begin{equation}
\ell_F^k = -\frac{1}{(T-W) N} \sum_{i=1}^{T-W} \sum_{n=1}^N [{y}^k_{i,n} \log (\hat{y}^k_{i,n}) + (1-{y}^k_{i,n}) \log (1-\hat{y}^k_{i,n})]. 
\end{equation}

There are two more design considerations. 

    \paragraph{Classification or Regression:} While formulating FBP as a regression task (predicting the frequency of each behavior) is possible, it may skew the model's focus towards more frequent events, overshadowing less common behaviors. Given the skewed behavior distributions in our data, we opt for a binary classification approach.
    
    
    \paragraph{Behaviors of Interest:} Which behaviors to predict depends on the application. For instance, a model aimed at ad click prediction might focus on ad-related behaviors. However, for broader downstream applications, we avoid manually selecting specific behaviors and instead include all possible behaviors for greater generalizability. 
    

\subsubsection{Same User Prediction}

The capability of discriminating different users is crucial for personalization, yet the FBP objective does not explicitly train the model for this. Thus, we introduce the same user prediction (SUP) objective. SUP encourages the model to assign similar embeddings to behavior sequences from the same user and dissimilar embeddings to behavior sequences from different users. We train the model with contrastive learning \citep{simclr}, which aims to increase the similarity between similar pairs of data while decreasing the similarity between dissimilar pairs. We randomly extract one pair of non-overlapping behavior sequences from each user to constitute positive samples and use in-batch negative sampling to obtain negative samples.

Let $\mathcal{X} = \{\bm{x}^k=[x_{1}^k,\ldots,x_{t^k}^k], \bm{x}^{k^+}=[x_{1}^{k^+},\ldots,x_{t^{k^+}}^{k^+}] \} $ define a batch of pairs of non-overlapping behavior sequences from the same user, where $M$ is the batch size, $t^k$ and $t^{k^+}$ are the sequence lengths. Let $\bm{e}^k$ denote the embedding of the behavior sequence $\bm{x}^k$. 
We adopt the SimCLR \citep{simclr} loss to implement SUP. Specifically, the loss function regarding anchor $\bm{x}^k$ is defined as follows:

\begin{equation}
\begin{split}
\ell^{k,k^+}_S =  
-\log \frac{\exp (\mathrm{sim}(\bm{e}^k, \bm{e}^{k^+}) / \tau)}
{\sum_{j \neq k} \exp(\mathrm{sim}(\bm{e}^k, \bm{e}^j) / \tau)} \;. 
\end{split}
\label{contrastive_loss}
\end{equation}
where $k$ and $k^+$ represent the indices of the anchor and the behavior sequence from the same user, $\tau$ is the temperature hyperparameter and $\mathrm{sim}(\bm{e}^k, \bm{e}^j)$ is the cosine similarity of $\bm{e}^k$ and $\bm{e}^j$. 
For every positive pair $(\bm{x}^k, \bm{x}^{k^+})$, we respectively take $\bm{x}^k$ and $\bm{x}^{k^+}$ as the contrastive anchor to calculate the contrastive loss. We also perform future $W$-behavior prediction on both $\bm{x}^k$ and $\bm{x}^{k^+}$.
Weighting both losses equally, the final loss function per batch is thereby:

\begin{equation}
    \mathcal{L} = \frac{1}{M} \sum_{k=1}^M \frac{1}{2} (\ell^{k,k^+}_S + \ell^{k^+,k}_S) + \ell_F^k + \ell_F^{k^+}
\end{equation}


\subsection{Implementation}

The USE model consists of $12$ Retention layers, $8$ retentive heads, a hidden size of $768$, and an intermediate size of $3072$. We set the future window size $W$ as $100$ for future $W$-behavior prediction (In Section~\ref{subsec:experiments_ablation}, we show the impact of choosing different $W$). To construct pairs of non-overlapping behavior sequences as inputs, we filtered out users with shorter than $1224$-length ($512$*2+$100$*2) behavior sequences. Each input behavior sequence has a sequence length of $512$, and we ensure a distance of at least $100$ behaviors between the pair of input behavior sequences to avoid information leakage. We train the model on $766130$ pairs of behavior sequences for $10$ epochs, with a global batch size of $512$ and a learning rate of $4e-4$. The learning rate linearly increases from $0$ to peak in the first $6$ percent steps and linearly decreases to 0 at the end. Training the user model takes about $36$ hours on $8$ NVIDIA V100 GPUs.

\section{Experiments}
In this section, we detail our experimental design and show empirical results. We focus on three research questions. 
\begin{itemize}
    \item \textbf{RQ1}: How do different training objectives impact the downstream performance of user embeddings in static settings? 
    \item \textbf{RQ2}: Does the stateful approach generate better embeddings than stateless approaches in dynamic settings?
    \item \textbf{RQ3}: How efficient is the stateful approach compared to stateless approaches in dynamic settings?
\end{itemize}

In static settings, user behavior sequences remain unchanged, whereas in dynamic settings, users continuously generate new behaviors and we periodically update user embeddings in response to the new user data, approximating real-world scenarios.

We describe our data and baseline models in Sections \ref{subsec:experiments_data} and \ref{subsec:experiments_baseline}, respectively. In Section \ref{subsec:experiments_static}, we present and discuss model performance on various tasks in static settings (RQ1). Finally, in Section \ref{subsec:experiments_dynamic} we delve into model performances in dynamic settings (RQ2\&3).

\subsection{Data}
\label{subsec:experiments_data}

We use user behavior sequences from Snapchat. 
To construct these sequences, we selected a total of $685$ distinct user behaviors, such as sending a chat, utilizing the camera, and applying a filter. We added a "new\_session" \textit{behavior} marker in each user's behavior sequence to signify the commencement of a session. An example sequence is: "new\_session, open\_app, open\_camera, read\_message, close\_app".

The dataset for training user models comprises of behavior sequences from a large, randomly selected sample\footnote{Note that we avoid reporting the actual sample sizes for all the analyses on purpose. However, we make sure that the sample sizes are up to industry and academic standards.} of U.S.-based adult users \textit{active} between April 1, 2023, and April 14, 2023. "Active" is defined as engaging with Snapchat for a cumulative duration exceeding one minute in this time frame. This sample was randomly divided into training and validation sets at a 19:1 ratio. In our downstream evaluations, to prevent information leakage, we only included behavior sequences from users who were not present in the user model training phase. All experiments were conducted across $3$ random seeds, and the results presented are the averages of these trials. 

\subsection{Baselines}
\label{subsec:experiments_baseline}
We compare USE with a range of baseline models.
\begin{itemize}
    \item Term Frequency (\textbf{TF}) and Term Frequency - Inverse Document Frequency (\textbf{TF-IDF}), which are traditional methods for vector representation.
    \item Skip-Gram with Negative Sampling (\textbf{SGNS}) \citep{mikolov_distributed_2013}, a model that learns a fixed vector for each user behavior by predicting context behaviors from a given target behavior.
    \item \textbf{Untrained} user representations, where each unique user behavior is represented by a randomly generated fixed vector. This approach has demonstrated competitive performance in various natural language tasks~\citep{arora-etal-2020-contextual}.
    \item Transformer Encoder trained with masked language modeling (\textbf{Trans-MLM}) and Transformer Decoder trained with causal language modeling (\textbf{Trans-CLM}). These are our implementations of BERT \citep{bert} and GPT2 \citep{gpt2} for behavior-based user modeling, utilizing architectures equivalent to \texttt{BERT-base} and \texttt{GPT2-117M}.
    \item Variants of USE, each trained with the same data and architecture but different training objectives: causal language modeling (\textbf{USE-CLM}), future $W$-behavior prediction (\textbf{USE-FBP}), and same user prediction (\textbf{USE-SUP})
\end{itemize}


TF and TF-IDF baselines generate fixed-length vectors representing entire behavior sequences, which can directly serve as user embeddings. Other models, such as SGNS, Untrained, BERT, and GPT2, learn vector representations for separate behaviors. Consequently, it is necessary to aggregate these behavior-level vectors into a sequence-level vector (i.e., a user embedding). For all models, we employ mean pooling to aggregate behavior vectors as user embeddings. Additionally, for autoregressive models like GPT2 and USE, we explored the effectiveness of using the last non-padded behavior's embedding as the user embedding. However, this method showed significantly inferior performance compared to mean pooling.

\subsection{Static User Modeling}

\begin{table}[t]
	\centering
         \caption{ 
        		Evaluation results on static user modeling, including $6$ tasks: User Retrieval (UR), Future Behavior Prediction (FBP), Reported Account Prediction (RAP), Locked Account Prediction (LAP), Ads View Time Prediction (ATP) and Account Self-deletion Prediction (ASP). We use bolded and underlined scores to indicate the best and second-best performances.
        	}\label{tb:static_results}
	\footnotesize
	\setlength{\tabcolsep}{2.4mm}{
	\begin{tabular}{lccccccc}\toprule
		
		\textbf{Model} & \textbf{UR}    & \textbf{FBP}  & \textbf{RAP} & \textbf{LAP} & \textbf{ATP} & \textbf{ASP} & \textbf{Ave.}\\

		\midrule

  {\textbf{TF} } & 15.4 & 78.7 & 89.0 & 93.1 & 88.7 & 58.1 & 70.5  \\

            {\textbf{TF-IDF} } & 17.5 & 78.9 & 88.3 & 92.5 & 88.0 & 60.4 & 70.9\\

            {\textbf{SGNS} } & 13.4 & 78.7 & 88.3 & 94.3 & 89.0 & 57.4 & 70.2 \\

            {\textbf{Untrained} } &15.2 & 78.9 & 89.0 & 94.0 & 88.8 & 59.9 & 71.0 \\

            {\textbf{Trans-MLM} } & 27.2 & 79.8 & \underline{90.6} & \textbf{95.3} & 89.7 & 62.8 & 74.2 \\

            {\textbf{Trans-CLM} } &33.2 & 79.8 & \textbf{90.7} & 94.4 & 90.0 & 63.8 & 75.3\\
		
		\midrule

            {\textbf{USE-CLM} } & 30.2 & 79.7 & 90.1 & 94.3 & 89.8 & 62.0 & 74.4\\

            {\textbf{USE-FBP} } & 36.1 & \textbf{80.3} & 90.0 & 94.0 & 89.5 & \textbf{64.7} & 75.8 \\

            {\textbf{USE-SUP} } & \underline{47.3} & 80.0 & 89.8 & 94.7 & \underline{90.0} & 63.9 & \underline{77.6} \\

            \midrule
            
            {\textbf{USE} (ours)} & \textbf{47.4} & \textbf{80.3} & 90.0 & \underline{95.0} & \textbf{90.4} & \underline{64.5} & \textbf{78.0}\\
            
		\bottomrule
	\end{tabular}}
\end{table}

\label{subsec:experiments_static}
In this section, we aim to answer RQ1. We evaluate model performance on the $6$ downstream tasks that utilize static user behavior sequences. In each task, user behavior sequences are fixed (i.e., not updated). For USE and each baseline, we compute user embeddings and use those as input for downstream evaluation.

\subsubsection{Evaluation Tasks}
\paragraph{User Retrieval (UR)}
This is a ranking task that evaluates a model's ability to distinguish between different users. The goal is to retrieve the correct user behavior sequence from a pool of 100 candidates, given a query behavior sequence. Each sample comprises 101 user behavior sequences: 1 \textit{query} and 100 \textit{candidates}. Among the candidates, there is 1 \textit{positive candidate} corresponding to the same user as the query and 99 \textit{negative candidates} belonging to other users. We represent each behavior sequence as a vector and rank the candidates based on their cosine similarity with the query. The model's performance is measured using Mean Reciprocal Rank (MRR): $\texttt{MRR} = \frac{1}{N} \sum_{i=1}^N \frac{1}{r_i}$, where $N$ is the number of samples and $r_i$ is the rank of the positive candidate in the $i$-th sample. Each behavior sequence contains 512 user behaviors. To increase task difficulty, negative samples are chosen from behavior sequences with a TF vector cosine similarity over $0.8$ with the query (hard negatives). 


\paragraph{Future Behavior Prediction (FBP)}
This multi-label classification task assesses a model's ability to predict future user behaviors. Specifically, it involves predicting whether a user will exhibit each behavior in their next 512 actions, akin to the future $W$-behavior prediction objective with a future window of 512. The dataset training, validation and testing ratio is 3:1:1. AUC is the evaluation metric.

\paragraph{Reported Account Prediction (RAP), Locked Account Prediction (LAP), Ads View Time Prediction (ATP), and Account Self-deletion Prediction (ASP)}
These four tasks involve predicting users who get reported by others, whose accounts get locked, who view an ad for a certain duration, and who delete their accounts voluntarily, respectively, on a given date (which we refer to as \textit{label date}). For each task, we vary the number of days (from 0 to 7) between the date of the last available user behavior and the label date. Here, $0$  corresponds to the users' behavior sequences that end on the label date but still before the timestamp of the event to predict.  Hence, for each task, we have eight evaluation datasets. Within each task, we evaluate the same set of users and ensure each dataset is balanced on class labels. Due to space limits, we only report the models' average performance across the 8 datasets in this section. The models' detailed performance on each task is presented in Appendix \ref{appendix:experiments_static}. We use AUC as the evaluation metric.

Except for user retrieval, all tasks involve training a single-layer MLP (multi-layer perception) classifier with user embeddings as input. Cross-validation with random search is employed for hyperparameter selection for the MLP classifier on each dataset. The input user behavior sequence length for all tasks is fixed at 512.

\begin{table*}[t]
	\centering
        \footnotesize
        \caption{ 
		Detailed evaluation results on Next-Period Behavior Prediction on 16 update periods of real-world simulation. $\clubsuit$ refers to the Recent Only strategy, which generates user embeddings purely based on new behavior sequences in each period.  $\spadesuit$ refers to the Pool Embeddings strategy, which independently computes an embedding at each period based on new user behaviors and averages all the embeddings as the user embedding. The $\diamondsuit$ refers to USE and Recompute All, which generate user embeddings conditioned on the entire user behavior sequence up to the current period. $\Delta$ represents the performance difference between $\diamondsuit$ and the second-best strategy.
	}
	\setlength{\tabcolsep}{1.3mm}{
	\begin{tabular}{lcrrrrrrrrrrrrrrrrr}\toprule

    \multirow{2}{*}{\textbf{Model}} &
		\multicolumn{16}{c}{\textbf{Performance on each update period}} & 
		 \\
		\cmidrule(lr){3-18} &
		& {\textbf{0}} & {\textbf{1}} &{\textbf{2}}  & {\textbf{3}} &{\textbf{4}} &{\textbf{5}} &{\textbf{6}} &{\textbf{7}} & {\textbf{8}} & {\textbf{9}} &{\textbf{10}}  & {\textbf{11}} &{\textbf{12}} &{\textbf{13}} &{\textbf{14}} &{\textbf{15}} & \textbf{Ave.} \\

		\midrule

             \multirow{2}{*}{\textbf{Trans-CLM}} &$\clubsuit$ & 68.8 & 68.8 & 69.1 & 68.9 & 68.8 & 69.1 & 68.2 & 69.6 & 69.3 & 69.0 & 69.2 & 68.7 & 68.8 & 69.0 & 68.9 & 68.7 & 68.93\\
             & $\spadesuit$ & 69.0 & 71.1 & 72.6 & 73.4 & 74.3 & 75.0 & 75.3 & 75.0 & 75.6 & 76.1 & 76.1 & 76.1 & 76.5 & 76.3 & 76.5 & 76.5 & 74.72\\

             \cmidrule(lr){1-19}

             \multirow{4}{*}{\textbf{USE-CLM}} & $\clubsuit$ & 69.2 & 70.1 & 70.1 & 70.3 & 70.7 & 70.3 & 70.1 & 70.8 & 70.4 & 71.0 & 70.6 & 71.0 & 71.1 & 70.7 & 70.3 & 70.8 & 70.46\\
             & $\spadesuit$ &69.4 & 71.7 & 73.2 & 73.6 & 74.2 & 74.5 & 74.5 & 74.7 & 75.0 & 75.4 & 75.4 & 75.6 & 75.8 & 76.0 & 75.8 & 75.8 & 74.41\\
             & $\diamondsuit$ & 69.4 & 71.7 & 73.3 & 73.8 & 74.7 & 75.1 & 75.0 & 75.3 & 75.7 & 76.4 & 76.2 & \underline{76.5} & 77.1 & \underline{77.0} & 76.8 & 76.9 & 75.06\\
             & $\Delta$ & 0.0 & 0.0 & 0.2 & 0.1 & 0.6 & 0.6 & 0.5 & 0.6 & 0.6 & 1.0 & 0.8 & 0.9 & 1.3 & 1.0 & 1.1 & 1.1 & 0.64\\

             \cmidrule(lr){1-19}

             \multirow{4}{*}{\textbf{USE-FBP}} & $\clubsuit$ & 69.5 & 70.1 & 70.6 & 70.3 & 70.1 & 70.6 & 69.7 & 70.5 & 70.7 & 70.7 & 71.4 & 70.7 & 70.7 & 70.7 & 70.5 & 70.5 & 70.46\\
             & $\spadesuit$ & 69.5 & 71.3 & 72.6 & 72.8 & 73.0 & 73.7 & 73.2 & 73.9 & 74.1 & 74.5 & 74.4 & 74.5 & 74.8 & 74.6 & 74.5 & 74.2 & 73.47\\
             & $\diamondsuit$ &69.5 & 71.4 & 72.7 & 73.0 & 73.6 & 74.5 & 74.3 & 75.0 & 75.1 & 75.8 & 75.8 & 76.0 & 76.3 & 76.1 & 76.2 & 76.0 & 74.46 \\
             & $\Delta$ & -0.0 & 0.1 & 0.1 & 0.2 & 0.7 & 0.8 & 1.2 & 1.1 & 1.1 & 1.3 & 1.4 & 1.5 & 1.5 & 1.5 & 1.7 & 1.8 & 0.99\\

            \cmidrule(lr){1-19}

             \multirow{4}{*}{\textbf{USE-SUP}} & $\clubsuit$ & 72.1 & 69.5 & 68.1 & 69.7 & 69.5 & 68.9 & 68.6 & 69.3 & 68.8 & 69.2 & 69.8 & 69.8 & 69.5 & 69.1 & 69.1 & 69.2 & 69.40\\
             & $\spadesuit$ & 72.1 & 73.0 & 73.8 & 73.7 & 73.7 & 73.9 & 74.1 & 74.0 & 74.1 & 74.2 & 74.2 & 74.5 & 74.7 & 74.2 & 74.5 & 73.9 & 73.92\\
             & $\diamondsuit$ & 72.1 & \underline{74.3} & \underline{76.0} & \underline{76.9} & \underline{77.2} & \underline{77.9} & \underline{77.7} & \textbf{78.3} & \underline{78.5} & \underline{78.6} & \textbf{78.8} & \textbf{79.1} & \textbf{79.3} & \textbf{78.9} & \textbf{79.0} & \underline{78.7} & \underline{77.59}\\
             & $\Delta$ & 0.0 & 1.3 & 2.2 & 3.2 & 3.6 & 4.0 & 3.6 & 4.3 & 4.4 & 4.4 & 4.6 & 4.6 & 4.6 & 4.8 & 4.5 & 4.8 & 3.67\\

            \cmidrule(lr){1-19}

             \multirow{4}{*}{\textbf{USE (ours)}} & $\clubsuit$ & \underline{72.4} & 70.3 & 70.4 & 70.8 & 70.7 & 70.6 & 70.4 & 70.6 & 70.4 & 70.3 & 71.5 & 71.0 & 70.7 & 70.5 & 71.0 & 70.4 & 70.74\\
             & $\spadesuit$ & \textbf{72.5} & 73.5 & 74.4 & 74.7 & 74.6 & 74.7 & 74.7 & 74.6 & 74.8 & 74.9 & 74.9 & 75.3 & 75.1 & 75.2 & 74.8 & 74.8 & 74.59\\
             & $\diamondsuit$ &\textbf{72.5} & \textbf{74.8} & \textbf{76.5} & \textbf{77.2} & \textbf{77.4} & \textbf{78.3} & \textbf{78.0} & \underline{78.2} & \textbf{78.6} & \textbf{78.7} & \underline{78.7} & \textbf{79.1} & \underline{79.1} & \textbf{78.9} & \underline{78.8} & \textbf{78.9} & \textbf{77.73}\\
             & $\Delta$ & 0.0 & 1.3 & 2.1 & 2.5 & 2.8 & 3.5 & 3.3 & 3.6 & 3.8 & 3.8 & 3.8 & 3.8 & 4.0 & 3.7 & 4.1 & 4.1 & 3.14\\

		\bottomrule
	\end{tabular}}
 \label{tb:next_period_prediction}
\end{table*}

\subsubsection{Results}

Table \ref{tb:static_results} shows the performance of different models across $6$ downstream tasks in static settings. These results suggest a significant advantage of sequence model-based methods over other baselines, underscoring the substantial potential and capability of sequence models in user modeling. Notably, Trans-CLM exhibits overall better performance compared to Trans-MLM, particularly in the User Retrieval (UR) task. This highlights the efficacy of causal language modeling in learning user embeddings, especially for tasks like UR. Despite employing the same training objective as Trans-CLM, USE-CLM demonstrates slightly inferior performance, suggesting a somewhat weaker representation capacity of RetNet compared to the Transformer architecture at this scale (i.e., 100M parameters). This observation aligns with findings in the original RetNet paper \citep{retnet}.
However, this slight decrease in representation capability is effectively offset by the adoption of more tailored training objectives. As the table indicates, both USE-FBP and USE-SUP outperform USE-CLM and Trans-CLM, attesting to the effectiveness of our proposed pre-training objectives. This effectiveness is further exemplified by the overall superior performance of our proposed method, USE. USE surpasses its individual objective-based variants, demonstrating the synergistic benefit of combining Future $W$-Behavior Prediction (FBP) and Same User Prediction (SUP) for the user model's downstream effectiveness.

\subsection{Dynamic User Modeling}
\label{subsec:experiments_dynamic}

In this section, we answer RQ2 and RQ3 by conducting simulations that approximate real-world scenarios where users continuously produce new behavior sequences, necessitating periodic updates to user embeddings to account for recent user behavior changes. We evaluate the effectiveness and efficiency of stateful user models in comparison to stateless models in such dynamic environments.



\begin{table*}[t]
	\centering
        \footnotesize
         \caption{ 
        		Evaluation results on User Re-Identification across 16 update periods of real-world simulation.  
        	}\label{tb:user_reid}
	\setlength{\tabcolsep}{1.3mm}{
	\begin{tabular}{lcrrrrrrrrrrrrrrrrr}\toprule

    \multirow{2}{*}{\textbf{Model}} &
		\multicolumn{16}{c}{\textbf{Performance on each update period}} & 
		 \\
		\cmidrule(lr){3-18} &
		& {\textbf{0}} & {\textbf{1}} &{\textbf{2}}  & {\textbf{3}} &{\textbf{4}} &{\textbf{5}} &{\textbf{6}} &{\textbf{7}} & {\textbf{8}} & {\textbf{9}} &{\textbf{10}}  & {\textbf{11}} &{\textbf{12}} &{\textbf{13}} &{\textbf{14}} &{\textbf{15}} & \textbf{Ave.} \\

		\midrule



             \multirow{2}{*}{\textbf{Trans-CLM}} & $\clubsuit$ & 29.4 & 28.5 & 27.6 & 27.7 & 26.8 & 27.1 & 26.4 & 25.8 & 25.7 & 25.4 & 25.4 & 25.5 & 25.0 & 24.7 & 24.0 & 24.4 & 26.21\\
             & $\spadesuit$ & 29.4 & 35.8 & 39.4 & 42.5 & 44.2 & 45.9 & 47.4 & 48.5 & 49.0 & 50.0 & 50.4 & 50.8 & 51.5 & 52.2 & 52.6 & 52.7 & 46.41\\

             \cmidrule(lr){1-19}

             \multirow{4}{*}{\textbf{USE-CLM}} & $\clubsuit$ & 27.0 & 32.4 & 32.7 & 33.1 & 32.7 & 33.2 & 32.1 & 31.3 & 31.6 & 31.5 & 31.4 & 31.1 & 30.7 & 30.2 & 29.9 & 29.9 & 31.29\\
             & $\spadesuit$ &27.0 & 32.4 & 35.5 & 38.4 & 40.2 & 42.1 & 43.4 & 44.6 & 45.3 & 45.8 & 46.2 & 46.9 & 47.5 & 48.4 & 48.8 & 49.2 & 42.60\\
             & $\diamondsuit$ & 27.0 & 35.5 & 39.6 & 43.6 & 45.8 & 48.0 & 49.3 & 50.4 & 51.3 & 51.9 & 52.3 & 52.8 & 53.5 & 54.5 & 54.9 & 55.1 & 47.84\\
             & $\Delta$ & 0.0 & 3.1 & 4.2 & 5.2 & 5.5 & 5.9 & 5.9 & 5.9 & 6.0 & 6.1 & 6.1 & 5.9 & 6.0 & 6.0 & 6.2 & 5.9 & 5.24\\

             \cmidrule(lr){1-19}

             \multirow{4}{*}{\textbf{USE-FBP}} & $\clubsuit$ & 26.4 & 40.1 & 44.4 & 46.4 & 46.1 & 45.3 & 44.9 & 44.4 & 43.9 & 44.0 & 43.3 & 43.9 & 43.1 & 42.5 & 42.1 & 42.2 & 42.68\\
             & $\spadesuit$ & 26.4 & 35.9 & 40.2 & 43.7 & 45.7 & 47.5 & 49.1 & 49.9 & 50.5 & 51.3 & 52.0 & 52.3 & 52.6 & 53.2 & 53.7 & 54.1 & 47.38\\
             & $\diamondsuit$ &26.4 & 37.7 & 44.0 & 48.5 & 51.6 & 54.0 & 55.7 & 57.0 & 57.9 & 58.8 & 59.1 & 59.7 & 60.3 & 60.8 & 61.1 & 61.3 & 53.35 \\
             & $\Delta$ & 0.0 & -2.4 & -0.4 & 2.1 & 5.5 & 6.4 & 6.6 & 7.2 & 7.4 & 7.5 & 7.2 & 7.4 & 7.6 & 7.5 & 7.4 & 7.1 & 5.25\\

            \cmidrule(lr){1-19}

             \multirow{4}{*}{\textbf{USE-SUP}} & $\clubsuit$ & \underline{34.0} & 36.0 & 35.1 & 34.8 & 33.4 & 33.7 & 33.6 & 33.3 & 32.2 & 32.0 & 32.0 & 33.3 & 32.1 & 32.0 & 31.5 & 32.0 & 33.18\\
             & $\spadesuit$ & \underline{34.0} & 44.8 & 50.3 & 53.2 & 54.4 & 56.3 & 57.6 & 58.4 & 59.2 & 59.6 & 60.0 & 60.5 & 60.7 & 61.0 & 61.2 & 61.4 & 55.78\\
             & $\diamondsuit$ & \underline{34.0} & \underline{46.0} & \underline{52.2} & \textbf{55.9} & \underline{57.6} & \textbf{59.2} & \textbf{60.5} & \textbf{61.5} & \textbf{62.3} & \textbf{62.9} & \textbf{63.0} & \textbf{63.4} & \textbf{63.9} & \textbf{63.8} & \textbf{63.8} & \textbf{64.0} & \textbf{58.36}\\
             & $\Delta$ & 0.0 & 1.2 & 1.9 & 2.7 & 3.1 & 2.8 & 2.9 & 3.1 & 3.0 & 3.3 & 3.0 & 3.0 & 3.1 & 2.8 & 2.6 & 2.5 & 2.58\\

            \cmidrule(lr){1-19}

             \multirow{4}{*}{\textbf{USE (ours)}} & $\clubsuit$ & \textbf{37.3} & 41.1 & 41.3 & 40.6 & 39.6 & 39.7 & 38.7 & 38.3 & 37.9 & 37.6 & 38.2 & 38.4 & 36.9 & 37.3 & 36.7 & 36.4 & 38.49\\
             & $\spadesuit$ & \textbf{37.3} & 46.0 & 51.5 & \underline{54.2} & 55.8 & 57.3 & 58.3 & 59.2 & 59.8 & 60.6 & 60.7 & 60.9 & 61.4 & 61.8 & 62.0 & 61.9 & 56.79\\
             & $\diamondsuit$ & \textbf{37.3} & \textbf{47.9} & \textbf{53.1} & \textbf{55.9} & \textbf{57.8} & \textbf{59.2} & \underline{60.0} & \underline{60.9} & \underline{61.6} & \underline{62.3} & \underline{62.2} & \underline{62.3} & \underline{62.7} & \underline{63.0} & \underline{63.3} & \underline{63.6} & \underline{58.31}\\
             & $\Delta$ & 0.0 & 1.9 & 1.6 & 1.6 & 2.0 & 1.8 & 1.7 & 1.7 & 1.8 & 1.7 & 1.5 & 1.4 & 1.4 & 1.1 & 1.4 & 1.7 & 1.52\\

		\bottomrule
	\end{tabular}}
\end{table*}

\subsubsection{Evaluation Tasks}
Our simulation uses behavior sequences of
a random sample of Snapchat users that are not present in the training dataset, with a span of $15$ periods. Initially, each user possesses a behavior sequence comprising $250$ behaviors. Subsequently, in each period, each user generates an additional $250$ behaviors (chosen based on the median number of behaviors per day for active users: $241$). At the end of each period, we update the user embeddings for downstream evaluation. We adapt the User Retrieval and Future Behavior Prediction tasks used in the static settings for the now dynamic settings. They are: \textbf{User Re-Identification} and \textbf{Next-Period Behavior Prediction}. The other four evaluation tasks in the static settings, however, cannot be implemented due to resource limitations (e.g., expensive queries). 


\textbf{User Re-Identification}: This task focuses on distinguishing users based on their behavioral patterns. We start by collecting a historical behavior sequence of $4000$ actions from each of the 
selected users and creating a corresponding historical embedding for each. During the simulation, user embeddings are updated at the end of each period. For each user, we rank all historical embeddings based on their cosine similarity with the current user embedding. Consistent with the User Retrieval task, Mean Reciprocal Rank (MRR) is employed as the evaluation metric.

\textbf{Next-Period Behavior Prediction}: This multi-label classification task involves predicting a user's probability of displaying specific behaviors in the subsequent period. We train an MLP classifier using the behavior sequences of an independent set of 
users. Throughout the simulation, user embeddings are updated at the end of each period. The updated embeddings are then used to make predictions via the trained MLP classifier. AUC serves as the evaluation metric.

\subsubsection{Effectiveness}

Table \ref{tb:next_period_prediction} and \ref{tb:user_reid} present model performance in our simulations, utilizing different strategies for computing user embeddings. The data in these tables consistently show that USE significantly outperforms the Recent Only and Pool Embeddings strategies across a range of settings. This underscores the importance of incorporating user history in the generation of user embeddings and highlights USE’s effectiveness in leveraging historical information. The Recent Only strategy, which entirely omits historical data,  yields the worst performance in nearly all scenarios. Predictably, it maintains a consistent performance level across different periods, given its reliance on a uniform amount of information for generating user embeddings.
In contrast, both Pool Embeddings and USE demonstrate better performance in later periods, benefiting from the accumulation of historical user data. Pool Embeddings shows a notable improvement over the Recent Only approach, indicating that even a simple average of user embeddings from different periods can significantly aid user modeling. However, it falls short of USE in almost every instance, and this performance gap widens with the progression of periods. This trend highlights the superiority of generating embeddings based on historical user states compared to independent embedding computations at each period.
Moreover, echoing findings from our static user modeling evaluations (see Section \ref{subsec:experiments_static}), USE surpasses baseline models in most settings. This further validates the effectiveness of our proposed training objectives and of stateful user modeling in dynamic settings.

\begin{wrapfigure}{l}{0.5\textwidth}
    \centering
    \begin{tikzpicture}
    \node[inner sep=0]{\includegraphics[width=0.5\columnwidth, trim={0cm 0.cm 0cm 0cm}, clip]{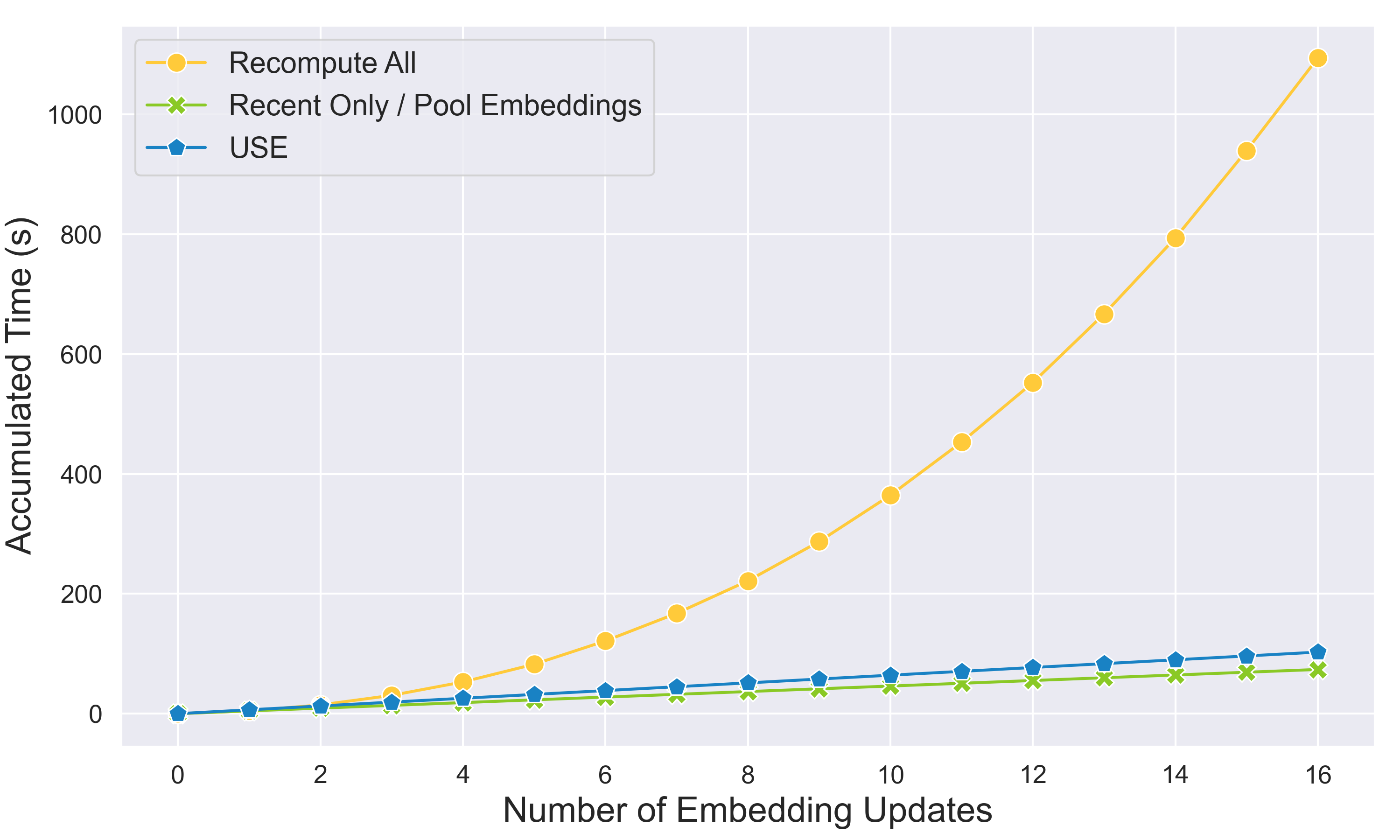}};
		\end{tikzpicture}
		\caption{Accumulated time costs of generating user embeddings in dynamic settings, across USE and three stateless strategies.}
    \label{fig:stateful_stateless}
\end{wrapfigure}

\subsubsection{Efficiency}
To fairly compare different methods' efficiency in generating embeddings in the dynamic setting, we compute the required cumulative time of each method for updating user embeddings at the end of each period in our simulation. Due to the varying memory usage with different methods, we dynamically adjust the batch size to saturate the GPU memory.
Figure \ref{fig:stateful_stateless} illustrates the required cumulative time of USE and $3$ stateless methods for updating user embeddings at the end of each period in our simulation. As the figure shows, USE demonstrates a consistent, constant time requirement for updates in each period, similar to the stateless methods that disregard historical data. In contrast, the Recompute All method incurs increasingly more time as the length of the user behavior history extends. The difference in efficiency between the stateful approach and the 'Recompute All' method is relatively modest at the beginning but becomes markedly significant over time. In real-world applications, where user behavior sequences can expand considerably, the stateful approach offers significant computational savings without compromising performance. Furthermore, while USE is slightly slower than the Recent Only and the Pool Embeddings approach, optimization of the USE implementation can help to minimize the efficiency difference.



\section{Conclusion}
In this work, we introduce the novel concept of stateful user modeling and conduct a comprehensive investigation, notably through the development and evaluation of our proposed Stateful User Embedding (USE) approach. Our experimental results demonstrate the significant advantages of USE in efficiently and effectively representing users in both static and dynamic settings. By leveraging the two training objectives of Future $W$-Behavior Prediction and Same User Prediction, USE not only addresses the limitations of traditional stateless models but also showcases its superiority in user representation. Our empirical evaluation using real-life behavior sequences from Snapchat users further confirms the effectiveness and efficiency of USE in generating user embeddings.

\paragraph{Broader Impact} We anticipate our proposed stateful user modeling approach to motivate a wider range of research, especially where the modeling targets dynamically evolve. For instance, our method can be readily applied to user modeling domains other than instant messaging apps (e.g., search engines, e-commerce websites). Moreover, this concept can apply to time-series analysis concerning dynamic targets, such as stock price and temperature forecasting, and conversational AI systems, which may store historical interactions with each user as user states for more personalized conversation.


\bibliographystyle{plainnat}  
\bibliography{references}

\newpage
\appendix

\section{Related Work}


\subsection{General-purpose User Modeling}
General-purpose user modeling is emerging as a powerful user modeling technique, offering adaptability to distinct downstream tasks such as item recommendations~\citep{chen_predictive_2018} and ad conversion prediction~\citep{wu_ptum_2020}, without the need to fine-tune the upstream user model. This stands in contrast to task-specific user modeling, where a separate user model is trained for each downstream task~\cite[e.g.,][]{modell_graph_2021, fan2019graph,waller_generalists_2019,zheng2017joint,liu2010personalized}), and thus incurs more computational costs.
There have been different approaches to general-purpose user modeling. For instance, Chen et al. (2018)~\citep{chen_forum_2018} computed user embeddings by aggregating user descriptive statistics on social media (e.g., minimum and maximum lengths of posts, user interest terms) in the form of vectors. 
Andrews and Bishop (2019)~\citep{andrews_learning_2019} used a Transformer-based encoder to learn user representations from the content and timestamps of users' social media posts. Sun et al. (2022)~\citep{sun2022learning} used a multi-interest constrastive learning algorithm to derive representations of both short-term and long-term user interests from user reviews and product ratings. 

\subsection{Behavioral Log-based User Modeling}
In addition to the approaches mentioned above, general-purpose user modeling based on user behavioral logs has become an increasingly popular and competitive alternative.
User behavioral logs are records of high-resolution, low-level events triggered by user actions in an information system~\citep{abb_reference_2022}.
General-purpose user modeling based on such data has demonstrated success in various domains.
For instance, Yang et al. (2017)~\citep{yang_personalizing_2017} used a word2vec algorithm to learn general user representations from behavioral sequences of Adobe Photoshop users, by predicting a specific user behavior given neighboring behaviors. 
Tao et al. (2019)~\citep{tao_log2intent_2019} modeled also Photoshop users based on their behavior sequences, however, with an encoder-decoder model (\textit{Log2Intent}). They defined two training objectives: predicting the next behavior and maximizing the semantic similarity between user behaviors and manual annotations of these behaviors. 
Chen et al. (2018)~\citep{chen_predictive_2018} used an RNN to learn representations for users of commercial websites based on their interaction sequences with the websites, by predicting the next behavior. 
More recent studies have adopted the Transformer architecture, enabling the model to learn contextual, more nuanced user representations.
For example, Zhang et al. (2020)~\citep{zhang_general-purpose_2020} modeled mobile phone users from their app usage sequences (e.g., app installation, uninstallation, retention, and timestamps) with two training objectives: reconstructing a user sequence and predicting masked behaviors. Chu et al. (2022)~\citep{chu_simcurl_2022} focused on professional design software, where they learned representations of users from their software command sequences, using a contrastive training objective. 
Pancha et al. (2022)~\citep{pancha_pinnerformer_2022} modeled Pinterest users based on their engagement sequences.
They propose a Dense All Action Prediction that encourages the hidden state of each randomly selected user behavior in the user behavior sequence to be similar to the Pins the user interacts with in the next $K$ days starting from the time this behavior. It is similar to our future $W$-behavior prediction with billions of candidate events. They solve it through contrastive learning with a negative sampling strategy.


\subsection{Stateless vs. Stateful User Modeling}
Most of the related work described above (except~\citep{chen_predictive_2018}) adopted a stateless user modeling paradigm, which does not compute user states. To update user embeddings, the model has to either discard (part of) historical user behaviors (which leads to less representation capacity) or recompute the entire user embedding with all the past and new behaviors (which is expensive). In contrast, stateful user modeling allows the storage and retrieval of previous user states, making dynamic updates of user embeddings natural and efficient. 
Given the almost non-existent work on stateful general-purpose user modeling, we review, instead, stateful but task-specific user modeling research below.

Examples of early approaches include temporal matrix factorization~\citep{koren2009collaborative}, Markov chains~\citep{he2016vista,he2016fusing,rendle2010factorizing} and RNNs~\citep{beutel2018latent,hidasi2018recurrent,liu2016context}. More recent approaches resort to memory-based networks~\citep{ren2019lifelong,pi2019practice}, where an explicit memory module is specified to learn and store information about users' past behaviors (i.e., user memories), updates these memories as new data comes in, and thus enables incremental updates of user embeddings over time. 
Our paper distinguishes itself from these earlier studies in three aspects.
First, we focus on general-purpose user modeling in a dynamic setting. 
Second, we adopt the Retentive Network architecture, which is not only Transformer-based (enabling greater representational capability and training parallelism, compared to the earlier approaches), but also RNN-based (allowing for low-cost inference and updates of user embeddings). 
Third, unlike earlier studies, we also comprehensively evaluate our approach in terms of both efficacy and efficiency compared to stateless approaches.

\section{Additional Experimental Results}

\subsection{Impact of $W$ in Future $W$-Behavior Prediction }
\label{subsec:experiments_ablation}

Thus, in this section, present empirical results on the impact of future window size $W$ used in the future $W$-behavior prediction objective. We train $5$ different models with only the future $W$-behavior prediction objective and evaluate them on the User Retrival task with different input lengths. Table \ref{tb:ablation} shows the results of models trained with different $W$. As shown in the table, models with a future window size of $50$ and $100$ achieves a similar level of performance, while the models performance drop greatly as $W$ increase. This is mainly because an extra-long window size results in mostly identical FBP labels (i.e., the existence of each unique behavior in the future $W$ behavior window) that encourage the model to predict almost the same target at each input behavior, which hinders the model from learning meaningful user engagement patterns.

\begin{table}[t]
	\centering
	\begin{tabular}{lrrrr}
		\toprule
  
		 \multirow{2}{*}{\textbf{$W$}} &  \multicolumn{4}{c}{\textbf{User Retrieval}} \\
    \cmidrule(lr){2-5}  & $512$ & $1024$ & $2048$ & $4096$ \\
   
		\midrule
		100 (ours)  & 29.98 & 40.12 & 49.29 & 58.67 \\
            50  & +0.31 & -0.06 & -0.26 & +0.08 \\
		200 & -1.54 & -1.22 & -0.64 & -0.85  \\
		500 &  -4.72 & -4.60 & -3.48 & -2.71\\
            1000 & -4.51 & -4.22 & -2.04 & -1.44\\
		\bottomrule
	\end{tabular}
	\caption{
		Ablation study on window size $W$ used in the future $W$-behavior prediction objective. We evaluate the models on User Retrieval task with different input lengths.
	}\label{tb:ablation}
\end{table}

\subsection{Static User Modeling}
\label{appendix:experiments_static}

\begin{table*}[ht]
	\centering
        \caption{ 
		Detailed evaluation results on Reported Account Prediction (RAP).
	}\label{tb:rap}
	\setlength{\tabcolsep}{2.4mm}{
	\begin{tabular}{lcccccccccc}\toprule

    \multirow{2}{*}{\textbf{Model}} &
		\multicolumn{8}{c}{\textbf{Number of days between input and label}} & 
		 \\
		\cmidrule(lr){2-9}
		& {\textbf{0}} & {\textbf{1}} &{\textbf{2}}  & {\textbf{3}} &{\textbf{4}} &{\textbf{5}} &{\textbf{6}} &{\textbf{7}} & \textbf{Ave.} \\

		\midrule

            {\textbf{TF} } & 91.50 & 89.30 & 89.10 & 89.10 & 88.60 & 87.60 & 88.30 & 88.10 & 88.95\\

            {\textbf{TF-IDF} } &90.50 & 89.00 & 88.80 & 88.10 & 87.60 & 88.40 & 87.20 & 87.10 & 88.34 \\
            
            {\textbf{SGNS} } & 91.90 & 88.60 & 87.30 & 88.50 & 88.30 & 88.20 & 87.60 & 86.10 & 88.31\\

            {\textbf{Untrained} } & 91.40 & 90.10 & 89.30 & 88.60 & 88.60 & 88.60 & 87.90 & 87.70 & 89.03\\

            {\textbf{Trans-MLM} } &93.10 & 90.80 & 90.80 & 91.10 & 89.40 & 90.00 & 90.20 & 89.40 & 90.60 \\

            {\textbf{Trans-CLM} } & 93.70 & 92.10 & 91.40 & 90.50 & 89.80 & 89.60 & 89.00 & 89.80 & 90.74\\
		
		\midrule

            {\textbf{USE-CLM} } & 93.50 & 90.20 & 90.00 & 90.00 & 90.50 & 89.20 & 88.10 & 89.00 & 90.06\\

            {\textbf{USE-FBP} } &92.70 & 89.70 & 90.50 & 89.90 & 89.60 & 88.70 & 90.30 & 88.20 & 89.95 \\

            {\textbf{USE-SUP} } & 92.50 & 90.50 & 89.50 & 89.50 & 89.80 & 88.70 & 89.10 & 89.10 & 89.84\\

            \midrule
            
            {\textbf{USE} (ours)} & 92.20 & 91.10 & 90.80 & 90.20 & 90.20 & 90.10 & 88.40 & 87.30 & 90.04\\
            
		\bottomrule
	\end{tabular}}
\end{table*}

\begin{table*}[ht]
	\centering
        \caption{ 
		Detailed evaluation results on Locked Account Prediction (LAP).
	}\label{tb:lap}
	\setlength{\tabcolsep}{2.4mm}{
	\begin{tabular}{lcccccccccc}\toprule

    \multirow{2}{*}{\textbf{Model}} &
		\multicolumn{8}{c}{\textbf{Number of days between input and label}} & 
		 \\
		\cmidrule(lr){2-9}
		& {\textbf{0}} & {\textbf{1}} &{\textbf{2}}  & {\textbf{3}} &{\textbf{4}} &{\textbf{5}} &{\textbf{6}} &{\textbf{7}} & \textbf{Ave.} \\

		\midrule

            {\textbf{TF} } & 97.40 & 95.60 & 93.30 & 92.10 & 90.30 & 91.60 & 91.80 & 93.00 & 93.14\\

            {\textbf{TF-IDF} } & 96.80 & 95.60 & 92.90 & 92.00 & 89.90 & 90.80 & 91.50 & 90.80 & 92.54\\
            
            {\textbf{SGNS} } & 98.00 & 96.60 & 94.70 & 93.60 & 91.80 & 93.00 & 93.00 & 93.60 & 94.29\\

            {\textbf{Untrained} } &97.90 & 96.30 & 93.80 & 93.40 & 92.50 & 92.80 & 93.30 & 91.90 & 93.99 \\

            {\textbf{Trans-MLM} } & 98.20 & 96.90 & 95.70 & 95.00 & 94.00 & 94.30 & 94.10 & 94.40 & 95.33\\

            {\textbf{Trans-CLM} } &98.30 & 97.20 & 94.70 & 94.10 & 93.00 & 93.20 & 91.80 & 93.20 & 94.44 \\
		
		\midrule

            {\textbf{USE-CLM} } & 98.00 & 95.80 & 94.80 & 94.80 & 92.80 & 93.40 & 92.70 & 92.30 & 94.33\\

            {\textbf{USE-FBP} } & 97.00 & 96.40 & 94.70 & 94.20 & 93.70 & 92.70 & 92.80 & 90.90 & 94.05\\

            {\textbf{USE-SUP} } &97.30 & 97.00 & 94.10 & 94.80 & 94.00 & 94.10 & 93.60 & 93.00 & 94.74 \\

            \midrule
            
            {\textbf{USE} (ours)} & 98.20 & 97.20 & 94.70 & 94.20 & 94.10 & 93.80 & 93.70 & 94.40 & 95.04\\
            
		\bottomrule
	\end{tabular}}
\end{table*}

\begin{table*}[ht]
	\centering
        \caption{ 
		Detailed evaluation results on Ads View Time Prediction (ATP).
	}\label{tb:atp}
	\setlength{\tabcolsep}{2.4mm}{
	\begin{tabular}{lcccccccccc}\toprule

    \multirow{2}{*}{\textbf{Model}} &
		\multicolumn{8}{c}{\textbf{Number of days between input and label}} & 
		 \\
		\cmidrule(lr){2-9}
		& {\textbf{0}} & {\textbf{1}} &{\textbf{2}}  & {\textbf{3}} &{\textbf{4}} &{\textbf{5}} &{\textbf{6}} &{\textbf{7}} & \textbf{Ave.} \\

		\midrule

            {\textbf{TF} } &96.90 & 89.50 & 89.20 & 86.60 & 86.50 & 86.80 & 85.90 & 88.00 & 88.67 \\

            {\textbf{TF-IDF} } & 95.60 & 88.80 & 88.60 & 85.80 & 85.50 & 87.00 & 86.90 & 85.40 & 87.95\\
            {\textbf{SGNS} } &97.40 & 89.80 & 88.60 & 87.30 & 86.40 & 87.30 & 87.40 & 87.50 & 88.96 \\

            {\textbf{Untrained} } &96.60 & 89.10 & 88.90 & 87.70 & 87.20 & 87.50 & 87.40 & 86.20 & 88.83 \\

            {\textbf{Trans-MLM} } &96.20 & 91.40 & 90.60 & 88.00 & 88.40 & 88.60 & 86.80 & 87.90 & 89.74 \\

            {\textbf{Trans-CLM} } & 97.70 & 91.00 & 90.10 & 89.20 & 88.20 & 88.60 & 87.60 & 87.70 & 90.01\\
		
		\midrule

            {\textbf{USE-CLM} } & 97.10 & 91.30 & 90.10 & 89.30 & 88.40 & 87.50 & 86.70 & 87.90 & 89.79\\

            {\textbf{USE-FBP} } &97.00 & 91.70 & 89.30 & 87.00 & 88.30 & 87.20 & 87.40 & 88.00 & 89.49 \\

            {\textbf{USE-SUP} } & 95.80 & 90.60 & 90.20 & 89.80 & 89.00 & 89.50 & 87.10 & 88.00 & 90.00\\

            \midrule
            
            {\textbf{USE} (ours)} & 96.20 & 91.80 & 90.70 & 89.40 & 88.70 & 89.80 & 88.20 & 88.30 & 90.39\\
            
		\bottomrule
	\end{tabular}}
\end{table*}

\begin{table*}[ht]
	\centering
        \caption{ 
		Detailed evaluation results on Account Self-deletion Prediction (ASP).
	}\label{tb:asp}
	\setlength{\tabcolsep}{2.4mm}{
	\begin{tabular}{lcccccccccc}\toprule

    \multirow{2}{*}{\textbf{Model}} &
		\multicolumn{8}{c}{\textbf{Number of days between input and label}} & 
		 \\
		\cmidrule(lr){2-9}
		& {\textbf{0}} & {\textbf{1}} &{\textbf{2}}  & {\textbf{3}} &{\textbf{4}} &{\textbf{5}} &{\textbf{6}} &{\textbf{7}} & \textbf{Ave.} \\

		\midrule

            {\textbf{TF} } & 63.20 & 55.30 & 62.10 & 62.50 & 65.10 & 50.50 & 54.30 & 52.00 & 58.13\\

            {\textbf{TF-IDF} } &63.10 & 51.80 & 63.10 & 58.00 & 56.80 & 63.70 & 63.80 & 62.90 & 60.40 \\
            {\textbf{SGNS} } &58.70 & 56.40 & 57.10 & 57.40 & 56.20 & 58.10 & 58.50 & 56.50 & 57.36 \\

            {\textbf{Untrained} } &60.00 & 61.00 & 59.00 & 61.90 & 59.00 & 59.90 & 60.00 & 58.40 & 59.90 \\

            {\textbf{Trans-MLM} } & 63.30 & 62.60 & 63.30 & 62.10 & 63.00 & 63.10 & 63.80 & 61.30 & 62.81\\

            {\textbf{Trans-CLM} } & 63.80 & 64.20 & 62.50 & 66.70 & 63.10 & 63.70 & 63.10 & 63.40 & 63.81\\
		
		\midrule

            {\textbf{USE-CLM} } & 62.80 & 61.30 & 61.90 & 61.70 & 61.30 & 66.40 & 60.80 & 60.10 & 62.04\\

            {\textbf{USE-FBP} } & 63.90 & 67.40 & 66.10 & 65.00 & 63.30 & 66.10 & 61.80 & 64.20 & 64.72\\

            {\textbf{USE-SUP} } &63.00 & 64.70 & 61.50 & 63.70 & 65.50 & 64.50 & 64.10 & 63.90 & 63.86 \\

            \midrule
            
            {\textbf{USE} (ours)} & 63.30 & 64.40 & 64.30 & 64.90 & 64.90 & 66.00 & 65.00 & 63.30 & 64.51\\
            
		\bottomrule
	\end{tabular}}
\end{table*}

In this section, we present additional experimental results on $4$ downstream tasks, Reported Account Prediction (RAP), Locked Account Prediction (LAP), Ads View Time Prediction (ATP), and Account Self-deletion Prediction (ASP). We collect $8$ datasets for each task. Each dataset contains the same users and labels, yet each of them contains behavior sequences of the users that are respectively 0/1/2/3/4/5/6/7 days before the label date, the date when the target event happens (e.g., a user was locked). As shown in the tables, USE achieves good performance across these tasks no matter the number of days between input and label. Moreover, in all the tasks, we observe a performance drop as the number of days between input and label increases from $0$ to $7$. However, in $3$ out of $4$ tasks, the models are able to achieve an AUC of over $0.8$ when the number of days between input and label is $7$, indicating the viability of accurately forecasting users' future engagements before at least one week.

\end{document}